Michael Perry[1], mperry20@gmu.edu
Hadi El-Amine[1], helamine@gmu.edu
[1] George Mason University, Department of Systems Engineering and Operations Research

# Computational Efficiency in Multivariate Adversarial Risk Analysis Models

**Abstract**

In this paper we address the computational feasibility of the class of decision theoretic models referred to as adversarial risk analyses (ARA). These are models where a decision must be made with consideration for how an intelligent adversary may behave and where the decision-making process of the adversary is unknown, and is elicited by analyzing the adversary's decision problem using priors on his utility function and beliefs. The motivation of this research was to develop a computational algorithm that can be applied across a broad range of ARA models; to the best of our knowledge, no such algorithm currently exists. Using a two-person sequential model, we incrementally increase the size of the model and develop a simulation-based approximation of the true optimum where an exact solution is computationally impractical. In particular, we begin with a relatively large decision space by considering a theoretically continuous space that must be discretized. Then, we incrementally increase the number of strategic objectives which causes the decision space to grow exponentially. The problem is exacerbated by the presence of an intelligent adversary who also must solve an exponentially large decision problem according to some unknown decision-making process. Nevertheless, using a stylized example that can be solved analytically we show that our algorithm not only solves large ARA models quickly but also accurately selects to the true optimal solution. Furthermore, the algorithm is sufficiently general that it can be applied to any ARA model with a large, yet finite, decision space.

## 1. Introduction

Decision making under uncertainty requires the consideration of probable outcomes; at the most rudimentary level a decision maker may assess a prior measure of risk such as $risk = (probability\ event\ occurs) \cdot (disutility\ of\ the\ event)$, and then make a decision with the intent of either reducing the probability of the event, decreasing disutility given the event occurs, or both. In certain disciplines such as engineering this simple approach may be acceptable. Complications arise, however, when the decisions you make are observed by an intelligent adversary whose actions are dependent on your own; in other words, $(probability\ event\ occurs)$ needs to be conditioned on your actions.

While the need to consider the effect of your actions on an adversary's behavior isn't a new idea, in practice the concept is often ignored even at the highest levels of national security (see, e.g. Brown and Cox (2011) and Hanley (2018)). Game theory provides a natural framework for addressing the problem of decision making in the face of an intelligent adversary and has evolved over the years to address the various shortcomings of its original conception, such as the optimistic assumption you know not only how you value various outcomes, but also how your adversary values outcomes. In brief, von Neumann is often cited as the founder of game theory due to his proof of the existence of a minimax solution in two-person zero sum games with perfect information (von Neumann, 1928); subsequent developments were made into solution concepts for more complex problems, the most pertinent in regards to the issue of adversarial uncertainty being the advent of Bayesian games (see Harsanyi (1967-68)). A more recent development for addressing adversarial uncertainty is a framework that has been termed adversarial risk analysis (ARA). In ARA, decision makers utilize traditional decision theoretic methodologies while treating their adversary's behavior like any other stochastic parameter, except that the distribution for these parameters is elicited through a prior belief on the distribution of the adversary's utility function and beliefs. General discussions on the merits of the decision theoretic approach to games can be found in Kadane and Larkey (1982) and Harsanyi, et al. (1982), as well as in Kadane, et al. (2011) and Banks, et al. (2016). ARA is a young field and the focus of this paper is to explore the computational feasibility of ARA models and develop a general-purpose heuristic algorithm for solving them.

In Section 3, the details of a general ARA model will be given and this model will serve as the basis for this paper. Each a decision maker and an adversary are expected utility maximizers and assumptions will be made about the decision maker's utility function, the adversary's stochastic utility function, and the probability distributions of various events, given courses of action of each the decision maker and adversary. Sensible assumptions will be made but the purpose of this paper is not to accurately estimate these quantities; our purpose is to assess the computational feasibility of ARA models as they grow in size and complexity.

In the majority of the ARA literature small examples are analyzed where decision makers have limited choices, and utility functions and densities are either binary or easily integrable. In general these conditions will not be met: decision makers typically need to make decisions that are multivariate; outcomes are often more nuanced than a simple binary measure of success or failure; and to accurately model the distribution of outcomes and utilities it would be ideal to have the full spectrum of functions available, regardless of whether they make integration easy. These factors cause exponential growth in the size of the model and to make things more challenging ARA requires the decision maker to assess her adversary's decision-making process, which will also typically be exponentially large. Making ARA models yet more challenging is the fact that the adversary's utility function and beliefs are assumed to be stochastic; this not only means an additional distribution must be integrated over, but there's also typically an elicitation process by which the decision maker assesses her adversary's preferences which involves a computationally expensive Monte Carlo simulation.

In light of these challenges, this paper addresses the problem of solving an ARA model with a continuous decision space where analytical solutions are unavailable, and a sufficiently fine discretization is used to solve for the optimal strategy. To the best of our knowledge no general-purpose algorithm has been suggested for handling ARA models of this form. Section 3 describes such a model, shows how to solve for the optimal decision in exact form (where we consider solutions based on a sufficiently fine discretization as exact), and gives descriptive formulae for the computational size of the exact model as a function of the dimensionality of the decision space. The problem blows up quickly, so Section 4 describes a simulation-based optimization technique to approximate the model in a feasible amount of time. Section 5 presents results of runtimes for variously sized example problems and assesses the statistical confidence in the accuracy our algorithm's output, and also compares our algorithm to an alternative methodology employed in the literature, where our methodology is seen to outperform. Section 6 provides an overview of future work that will complement the methodologies described in Section 4.

## 2. Literature review

Rios and Rios Insua (2012) lay out mathematical formulations and give numerical examples for three simple versions of ARA that can be thought of as building blocks for larger models: (i) a simultaneous defend-attack model; (ii) a sequential defend-attack-mitigate model; and (iii) a sequential defend-attack model with private defender information. Various ways their methodology can be expanded to model more detailed scenarios are discussed by Banks, et al. (2016) in a comprehensive monograph on ARA. These features include alternative methods to account for level-k thinking (i.e. "he thinks, that I think, that he thinks, that I think …"), multiple adversaries and/or allies, and the expansion of ARA to complex systems that can't be modeled as simple sequences of actions.

In the existing literature on ARA there are a multitude of "toy" examples that have been used to illustrate the above and other concepts while keeping the computational effort low. For example, optimal military convoy routing under the threat of improvised explosive devices (IEDs) was modeled by Wang and Banks (2011). Many researchers have studied the optimal allocation of counterterror resources; see, e.g., Rios and Rios Insua (2012), Sevillano, et al. (2012), and McLay, et al. (2012). Optimal bidding in auctions was studied in Rios Insua, et al. (2009). Looking beyond these toy models, the most extensive example we found in the literature was in Banks, Rios, and Rios Insua (2016) and is based on an actual ARA performed for a client that operates a railway system with multiple stations and is concerned with pickpockets and fare evaders. While the client's decision space is quite large, representing possible security measures, the adversarial decision space is limited (to steal or not to steal), so the problem size remains relatively small. Nevertheless, an exact solution for the optimal security portfolio was computationally impractical so the analysts employed a greedy algorithm with random restarts to search for local optima; in Section 5 we'll utilize their approach in our example described in Section 3 and show that our methodology outperforms.

As noted in Section 1, ARA is just one approach to analyzing problems of adversarial conflict. Game theorists in general have studied the above listed applications and as the field evolved analyses accounting for uncertainty in the behavior of other players has become the norm, not the exception. Allocating resources against terrorism and other criminal activities has been studied in Wang and Bier (2011), Nikoofal and Zhuang (2012), and Liang and Xiao (2013), to name a few. Protection from IEDs was studied using traditional game theory by Lin and Dayton (2011). Game theory is ripe with other applications that have not yet been analyzed from an ARA perspective and make for promising future ARA studies; these include models of conventional warfare (e.g. Kovenock and Roberson (2012)), negotiations among political actors to include fringe figures such as moderate terrorists (e.g. Bueno de Mesquita (2005), Lapan and Sandler (1988), Bueno de Mesquita (1997)), and coordination games among willing participants in revolutionary activities (e.g. Kiss, et al. (2017), Chwe (2000), Edmond (2013)). While the above applications would be interesting to analyze using ARA, meaningful conclusions can only be drawn if methodologies for solving reasonably sized problems quickly are developed; literature on computational feasibility is where we direct our attention now.

In their book, Banks, et al. (2016) acknowledge computational feasibility will likely be an issue as ARA is developed in greater detail, but that the problem is similar to that faced by other decision theory problems. An example of solving game theory problems with many decision makers and complex sequences of decision can be found in Koller and Milch (2003), where the notion of strategic dominance is used to decompose a problem into several smaller problems that can be solved in sequence. Even in the smaller subproblems the decision space can become exponentially large for the reasons discussed in Section 1. If the decision space is a continuous random variable it may be possible to derive analytic solutions to ARA and game theory models, as was done in Zhuang and Bier (2007). However, such results usually rely on closed form expressions for the derivatives of the players' expected utility, which in general will not be possible. We make the assumption in this paper the decision space must be discretized and thus consists of a large yet finite set of possible decisions, each of which is subject to an uncertain outcome due to uncertainty about how an adversary will behave. Statistical selection methods that attempt to select the best among many alternatives have been used extensively in non-adversarial decision problems with uncertainty, and generally seek to optimize a measure of confidence the selected alternative is indeed the true optimum. These methods are often grouped by their overarching methodology; popular methodologies include optimal computational budget allocation (OCBA), the expected value of information (e.g. Chick, et al. (2010)), and sequential elimination methods (e.g., Fan, et al. (2016)). The OCBA methodology first developed by Chen, et al. (2000) will be used as the overarching framework in this paper. OCBA has been further developed since its initial conception to account for nuanced, problem-specific applications, and the state-of-the-art methods of OCBA as well as evidence of its sustained superior performance can be found in the textbook by Chen and Lee (2011), and a recent paper by Peng, et al. (2016).

While powerful, OCBA still requires a preliminary assessment of *all* possible decisions, and thus also may become impractical. Methods combining partitioned based search to identify promising regions of the decision space, and OCBA within the most promising regions, have been explored by, e.g., Chen, et al. (2014) and Xu, et al. (2016). Another method that can be used to identify promising regions of a decision space is that of Bielza, et al. (1999), who define an artificial probability distribution for expected utility using the cross section of decision and state variables, which is calibrated via Markov Chain Monte Carlo methods. Determining the marginal distribution in terms of only the decision variables then indicates where in the decision space the optimal decision lies.

## 3. Model formulation

*3.i. The model*

For our analysis we'll use a two-player sequential game between a decision maker (referred to as "she" throughout this paper) and her adversary ("he"), where the adversary acts second. As in all ARA models we take the perspective of the decision maker and thus her beliefs and utility function are known to us. In this model, the decision maker implements a strategy, the adversary fully observes it, and then implements his response strategy. Sequential models that continue for multiple moves are obvious extensions of this. Simultaneous models can also be considered an extension, as simultaneous games are often solved using the concept of level-k thinking: a level-1 thinker responds to the expected actions of an actor who doesn't consider his adversary's actions, a level-2 thinker responds to the expected actions of a level-1 thinker, and so on. Players in a simultaneous game are therefore basing their decisions on the expected actions of their adversary, which is mathematically identical to the first mover's problem in a sequential game.

We consider the situation where the decision maker and adversary are competing over $n$ distinct targets, which could represent anything from tactical-level concerns, such as securing a command center from attack, to high-level strategic concerns, such as some measure of economic control over a region. A strategy of the decision maker is denoted as an $n$-dimensional vector, $d$, where each element of $d$ represents a particular target towards which she may allocate resources. Similarly, an adversarial strategy $a$ is an $n$-dimensional vector stating how much resources the adversary invests towards each target. For each target there will be a stochastic level of success achieved, with 1 representing the adversary's most favorable outcome and 0 the decision maker's. Denote these levels of success as $S \in R^n$. Given these definitions we're now ready to describe the ARA model.

**Figure 1: decision trees for the sequential game**

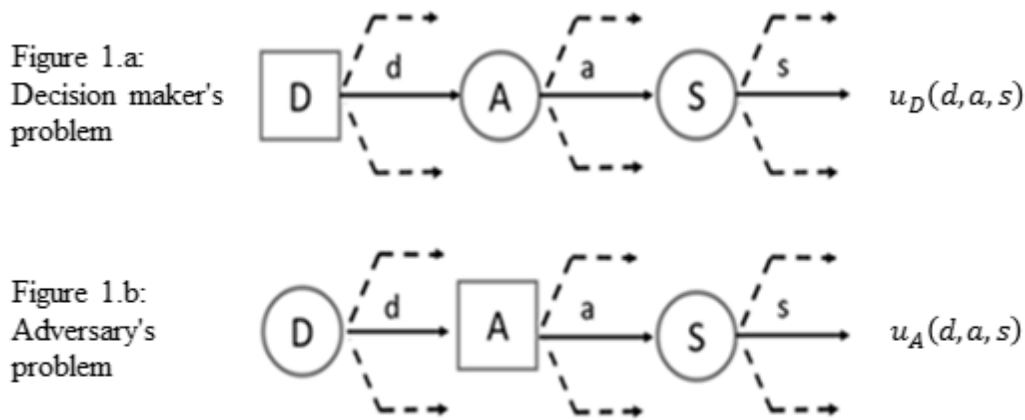

Figure 1.a: Decision maker's problem

Figure 1.b: Adversary's problem

Figures 1.a and 1.b represents the decisions faced by each the decision maker and adversary, respectively. Circles represent chance nodes where he or she must calculate expected utility given the decisions made up to that point, and squares represent decision nodes. As seen in figure 1.a, the defender can solve her problem of maximizing utility as follows:

**D1. For each $(d, a)$, find the decision maker's expected utility at chance node $S$:**

$\psi_D(d, a) = \int_S u_D(d, a, s) \cdot p_D(s|d, a) ds$, where $u_D$ is the decision maker's utility function and $p_D(s|d, a)$ is her belief about the density of $S$, given $d$ and $a$.

**D2. For each $d$, find the decision maker's expected utility at chance node $A$:**

$\psi_D(d) = \int_A \psi_D(d, a) \cdot p_D(a|d) da$, where $A$ is a random variable representing the adversary's strategy given $d$, and $p_D(a|d)$ is the decision maker's belief about its density.

**D3. Maximize expected utility at decision node $D$:**

$d^* = \underset{d \in F_D}{\mathrm{argmax}}\, \psi_D(d)$, where $F_D$ is the set of all feasible strategies.

Recall the fundamental principle of ARA that differentiates it from standard game theory: the adversary's utility function and beliefs are uncertain and are elicited using the decision maker's beliefs about her adversary. While the adversary's utility function and density of $S$ are not explicitly present in steps D1 through D3, they materialize implicitly through the decision maker's ability to assess the distribution of $a$, $p_D(a|d)$, in step D2. Without perfect knowledge of these quantities she's uncertain how her adversary will behave and hence $p_D(a|d)$ is unknown. To overcome this challenge she must make an assumption on the functional form of the attacker's utility function, $U_A(d, a, s, r_u)$, and density of $S$, $P_A(s|d, a, r_p)$, where $r_u$ and $r_p$ are (possibly multidimensional) realizations of a random variable $R = \begin{bmatrix} R_u & R_p \end{bmatrix}$ governing the specific forms of $U_A$ and $P_A$. Also note we've used capital letters to emphasize these are stochastic functions on account of uncertainty in $R$; this convention will be used throughout the paper. Making an assumption on the distribution of $R$, the decision maker can now infer a distribution on her adversary's behavior, $P_D(a|d)$, by analyzing the problem from his perspective. For each possible value of $d$, we trace figure 1.b backwards from node S to A in the following manner:

**A1. Randomly sample a value of R:**

Using the assumed distribution of $R$, the decision maker can randomly sample a single value. Denote this as $r^i = \begin{bmatrix} r_u^i & r_p^i \end{bmatrix}$.

**A2. For a given $d$, for each $a$ find the adversary's expected utility at chance node $S$:**

$$\Psi_A^i(d, a) = \int_S U_A(d, a, s, r_u^i) \cdot P_A(s|d, a, r_p^i) ds$$

**A3. Generate a sample of the optimal adversarial strategy at node A, given $d$:**

$a^i(d) = \underset{a \in F_A}{\operatorname{argmax}} \, \Psi_A^i(d, a)$, where $F_A$ is the set of all feasible adversarial strategies.

**A4. Infer the density $p_D(a|d)$:**

Repeating steps A1 - A3 $N_R$ times generates equally likely samples, $a^1(d), a^2(d), ...,$ $a^{N_R}(d)$, for the optimal adversarial strategy, given $d$. These samples serve as an estimate the decision maker can use for the density of the adversary's actions, denoted $P_D(a|d)$.

Repeating this process for all $d$ gives inferred densities that can be used in place of $p_D(a|d)$ in step D2, and so the decision maker can now solve her utility maximization problem.

*3.ii. Size of the model*

Analyzing the above model reveals three factors that affect its computational size: the number of feasible strategies, the number of samples of $R$ drawn in step A4, and the precision with which you'll estimate integrals over $S$, under the assumption the functional forms of $u_D, U_A, p_D(s)$, $P_A(s)$, and $P_D(a)$ don't offer an analytical solution. The appropriate number of samples to draw from $R$ will depend on its dimensionality, so we'll assume it contains $n$ elements of uncertainty;

recall that $n$ is the number of targets being competed over. Denote the number of samples to be drawn in A4 as $N_R$, and denote the number of intervals to use when discretizing the integrals over each element of $S$ as $N_S$. We will fix the values $N_R = 10^n$ and $N_S = 10$ and proceed to analyze the size of the problem as the number of feasible strategies increases, as this is likely to be the most pertinent parameter to a strategic planner using an ARA model.

The number of feasible strategies will be influenced by two factors: *(1)* the dimensionality of $d$ and $a$ (i.e. $n$); and, under the assumption the decision space is in reality continuous but must be discretized for computation, *(2)* the discretization used for $d$ and $a$. We will assume it's sufficient for each element of $d$ and $a$ to take on values in the set $\{0, .1, .2, ..., .9, 1\}$. The only parameter influencing problem size not yet fixed is $n$, and analyzing the computational feasibility as $n$ increases will make up the remainder of this paper.

As a preliminary analysis note that as $n$ increases from 2, to 3, 4, 5, the number of feasible strategies for the decision maker, $|F_D| = \binom{9+n}{n-1}$, increases from 11, to 66, to 286, to 1,001. An identical result obviously holds for $|F_A|$. It can also be shown that the number of integrals over $S$ that must be performed to solve the ARA model is:

$$number\ of\ integrals = |F_D| \cdot |F_A| \cdot (1 + N_R)$$

Now, considering that integrating over $S$ is an $n$-fold integral we're estimating numerically using $N_S$ intervals per element of $S$, the actual number of computations required to solve the ARA model is:

$$number\ of\ computations = |F_D| \cdot |F_A| \cdot (1 + N_R) \cdot N_S^n$$

In this paper we'll examine 4 cases for the values of $n$: $n \in \{2,3,4,5\}$. The growth in the number of computations in $n$ is enormous, increasing from over 1 million when $n = 2$, to over 4 billion when $n = 3$, 8 trillion when $n = 4$, and finally 10 quadrillion when $n = 5$. When none of the requisite integrals have analytical solutions the model becomes computationally intractable in its exact form when $n = 3$, so as discussed in Sections 4 and 5 approximation techniques must be used to solve larger problems. Section 4 discusses the methods actually employed by us and take the approach of analyzing the whole decision space in a computationally efficient way. Section 6, where future work is discussed, addresses approaches for finding promising regions of the decision space; if identifying these regions can compress the decision space of extremely large problems, then the methods of Section 4 can be used to thoroughly assess these relatively small spaces.

## 4. Methodology

The simulation-based methodology that will be used for solving the ARA model of Section 3 is based on OCBA, which is a statistical selection technique that's been employed in a variety of contexts to find good solutions quickly. We'll compare both the solve times of our statistical selection method to those of the exact model, as well as our method's ability to converge to the true optimum. To facilitate the latter we'll select utility functions, $u_D(d, a, s)$ and $U_A(d, a, s, r_u)$, and the distributions $p_D(s|d, a)$ and $P_A(s|d, a, r_p)$ so the integrals in steps D1 and A2 give analytical solutions for $\psi_D(d, a)$ and $\Psi_A(d, a)$ and the optimization over the adversary's behavior in A3 can be solved efficiently using standard algorithms, and thus the model can be solved in exact form in a reasonable amount of time. Because in practice we're interested in solving problems without analytical solutions for $\psi_D(d, a)$ and $\Psi_A(d, a)$, and where the optimization for the adversary's behavior is hard, we'll also attempt to solve the model in exact form by discretizing these integrals and then solving the optimization model in A3 via full enumeration for the purpose of assessing the computational advantages of statistical selection.

Before discussing the details of how we employ OCBA we present a narrative example for the ARA model to be solved, as well as the functional forms of its utility and density functions.

*4.i. A modified Colonel Blotto game*

Colonel Blotto games are a class of widely studied game theory models where one military commander, Colonel Blotto, must allocate her forces to $n$ distinct battlefields, knowing that her adversary, Colonel Klink, is simultaneously making the same decision. The objective is to win control of the most battlefields where each field is won based on some (possibly stochastic) function of the amount of forces deployed. We modify this example to assume there's an ongoing military conflict in which Colonel Blotto is considering intervening. If she intervenes to aid her allies, she knows with certainty that Colonel Klink will respond by intervening in aid of the other side, but she's uncertain as to which battlefields he values most and hence how he'll respond. We assume Blotto's only uncertainty about Klink is in his utility function, and that Blotto and Klink share a common density over the outcome on a battlefield, $p(s|d, a) = p_D(s|d, a) = p_A(s|d, a)$. Note also we've modified the game to be a sequential model: Colonel Klink only acts after observing Colonel Blotto's decision.

In this context, Colonel Blotto is our decision maker and Colonel Klink the adversary. The decision vectors $d$ and $a$ now represent what percentage of forces each Blotto and Klink deploy to each of the $n$ zones, and $S$ is a random vector representing the degree of success in each zone ($S_i \in [0,1]$, where $S_i = 1$ is the best possible outcome for Colonel Klink).

### 4.ii. *Utility and density functions*

We assume the levels of success on each battlefield follow independent uniform distributions over intervals of length 0.1, $S_i \sim Uni(h_i, h_i + .1)$, where $h_i \in [0, .9]$ depends on the $i^{th}$ elements of $d$ and $a$. In particular, $h_i(d_i, a_i) = -\frac{2}{4.6}\left(\log(c_{A,i} a_i + 1) - \log(c_{D,i} d_i + 1)\right) + C_{H,i}$, where $c_{A,i}, c_{D,i}$, and $C_{H,i}$ are measures of how difficult it is to attack target $i$. Blotto's and Klink's utility functions are:

$$u_D(s) = \frac{\sum_{i=1}^{n} v_i \left(e^{-4.6(s_i - C_{H,i} - .05)} - 1\right)}{n},$$

$$U_A(s, r) = \frac{\sum_{i=1}^{n} r_i \left(1 - e^{-4.6(s_i - C_{H,i} - .05)}\right)}{n},$$

where $v_i$ is the known value Blotto places on battlefield $i$, and $r_i$ is the uncertain value Klink places on battlefield $i$.

The proof that these specifications result in analytic solutions for $\psi_D(d, a)$ and $\Psi_A(d, a)$ is in Appendix A1, and the formulation for $a^i(d) = \underset{a \in F_A}{\operatorname{argmax}} \Psi_A^i(d, a)$ as a simple quadratic integer program is in Appendix A2. The intuitive appeal of these specifications is that $h_i$ is increasing in $a_i$ and decreasing in $d_i$, and determines a relatively tight bound of length 0.1 on the possible outcomes of each $S_i$ based on how much effort each Blotto and Klink apply to field $i$. The rate of change of $S_i$ with respect to $a_i$ and $d_i$ is driven by the (fixed) parameters $c_{A,i}$ and $c_{D,i}$. If each Blotto and Klink invest zero resources on zone $i$, then $h_i = C_{H,i}$. In the example we analyze in Section 5, $C_{H,i}$ will be non-zero, indicative of the fact they're intervening in an ongoing conflict where, without intervention, the $S_i$ values will be non-zero. The utility functions are exponential, and if all $S_i = C_{H,i} + .05$, representing no change to the status quo, then utilities evaluate to zero. Values of $S_i$ above $C_{H,i} + .05$ lead to positive utility for Klink proportional to the importance he attaches to battlefield $i$, $r_i$, while values below $C_{H,i} + .05$ lead to positive utilities for Blotto proportional to $v_i$. The "4.6" constant was chosen so that in the extreme case when $C_{H,i} = 0$ and $s_i = 1$, Klink receives utility of approximately $\frac{r_i}{n}$ and Blotto is decremented by $\frac{v_i}{n}$ (because $e^{-4.6} \approx 0$). For the distribution of the unknown parameters $R_i$ of which $r_i$ are realizations, we assume each $R_i$ follows a triangular distribution: $R_i \sim triangular(l_i, m_i, u_i)$. In practice, these model parameters would be fit rigorously by codifying intelligence analysts' knowledge into measurable functions and distributions.

### *4.iii. Optimal Computational Budget Allocation for solving large ARA models*

To solve for Colonel Blotto's optimal decision we'll utilize an algorithm based on the Optimal Computational Budget Allocation (OCBA) scheme of Chen, et al. (2000). OCBA is a general technique to support decision making under uncertainty. While it assumes normality of $u_D$, theoretical methods that generalize OCBA to any distribution are a current point of research and a few non-normal extensions have been derived in Glynn and Juneja (2004), and Chen, et al. (2000) also showed that in practice OCBA performs well for non-normal utility functions. Empirical results for the example presented in this paper show highly normal behavior (see figure 3 of Section 5.ii).

Developed for the purpose of choosing a good strategy under uncertainty when fully evaluating all possibilities is impractical, OCBA evaluates potential strategies through a series of iterations. In the first iteration all strategies are given preliminary examination by taking a small number of samples of the random variables causing the uncertainty (in our case, $S$ and $R$). In subsequent iterations, OCBA solves a nonlinear optimization problem to maximize an approximation of the probability of selecting the best strategy, subject to a constraint on the total number of samples to be drawn per iteration, and the process continues until either sufficient confidence an optimal strategy has been found is achieved, or the total computational budget is exhausted. Formally, during each iteration OCBA's sampling allocation scheme yields the result in theorem 1:

**Theorem 1 (proved in Chen, et al. (2000))**
Given a finite number of samples to be allocated among $k$ competing strategies, whose utility each follow a normal distribution, $u_i \sim N(\mu_i, \sigma_i^2)$ for $i = 1, 2, \ldots, k$, the approximate probability of funding the true optimal strategy (APCS) is asymptotically maximized using the following allocation scheme:

$$\frac{N_i}{N_j} = \left( \frac{\sigma_i/(\mu_b - \mu_i)}{\sigma_j/(\mu_b - \mu_j)} \right)^2, i \neq j \neq b, \quad (1)$$

$$N_b = \sigma_b \sqrt{\sum_{i \neq b} \frac{N_i^2}{\sigma_i^2}},$$

$\sum_{i=1}^{k} N_i =$ total sampling budget, and
$b = \underset{i}{\mathrm{argmax}}\, \mu_i$ (i.e. the index for the best strategy, based on the highest mean utility).

When applying theorem 1 in practice, the normal parameters $\mu$ and $\sigma$ are estimated using the most current sample data after each iteration of the OCBA-based algorithm described above. The approximation for the probability of correctly selecting the true optimum, APCS, is the objective function being maximized and is defined using the Bonferonni (1936) lower bound on the actual probability of correct selection:

$$P(u_b > u_i, \forall\, i \neq b) \geq 1 - \sum_{i \neq b} P(u_i > u_b) = 1 - \sum_{i \neq b} \Phi \left( \frac{\mu_i - \mu_b}{\sqrt{\frac{\sigma_b^2}{N_b} + \frac{\sigma_i^2}{N_i}}} \right) := APCS, \quad (2)$$

where $\Phi(\cdot)$ is the standard normal cumulative distribution function.

The choice of APCS as the metric to be maximized is partly pragmatic, as is allows the optimal sampling problem in theorem 1 to be solved analytically. However, APCS is also an appealing metric from a decision theoretic standpoint as it's a lower bound on the true probability of correct selection (PCS); lacking a reliable metric for PCS, a decision maker will likely be most interested in knowing how low PCS could be (i.e. APCS). Also, as shown in Chen and Lee (2011), APCS converges to PCS under mild conditions and numerical tests show strong performance in identifying strategies with the true PCS. Aside from fixing a total computational budget across iterations, there's no general-purpose stopping criterion that's used in the OCBA literature. A sufficiently high APCS may seem a natural stopping criterion but has issues when two strategies have expected utilities that are close. In our ex post analysis, we outline a simple method for adjusting APCS when two or more expected utilities are nearly indistinguishable, but for a stopping criterion we instead define a criterion based a lack of material changes in the most promising strategies from one iteration to the next; this is detailed in Appendix A3.

Applying this to our ARA model, for each $d \in F_D$ we need to draw samples of Colonel Blotto's objective function, $u_D(d, a, s)$, for all values of $d$. This is done by simultaneously drawing samples of $S$ and $A$. Sampling a value $S = s$ is straightforward as each element is $S_i \sim Uni(h_i, h_i + .1)$, with $h_i$ deterministically determined by $d_i$ and $a_i$. The difficulty is in sampling $A = a$ as $p_D(a|d)$ is unknown. To overcome this, we sample instead a value $R = r$ and then transform this into a sample of Colonel Klink's optimal behavior by solving $a^*(d) = argmax_a \int_S U_A(d, a, s, r) \cdot p(s|d, a) ds$. Because this is an $n$-dimensional integral that may not have an analytical solution, it becomes computationally expensive as $n$ increases. Thus, we'll perform a nested OCBA to evaluate it, where for each $a \in F_A$ samples from $S$ are drawn according to the OCBA sampling allocation rules. Together with the sample $s$ in the higher-level OCBA representing Blotto's problem, this gives a single sample of $u_D(d, a, s)$. Repeating the process per the allocation scheme in (1) gives a sample set that can be used to estimate $\psi_D(d)$; repeating for all $d$ allows you to compare strategies and recompute allocation scheme (1) for the next iteration.

Algorithm 1 formalizes the solution technique to be used in Section 5. To provide a descriptive summary of the algorithm, for each strategy $d$, multiple draws are taken from $(S, R)$ and each of these draws requires performing a nested OCBA to transform the sampled value of $R$ into a sampled value of $A$. The drawn samples are used to estimate the utility of each strategy $d$, and the process repeats until the stopping criteria in Appendix A3 is met. We refer to this process as one "trial," as in a Bernoulli trial where a particular strategy either is or is not proposed as the optimal strategy. We then continue to run trials until we are 95% confident the most frequently observed outcome of a single trial is in fact the most likely outcome, as calculated using Wilson's (1927) Bernoulli confidence interval:

> **Lemma 1 (see proof in Appendix A4)**
> Assume an experiment is being conducted with $k$ possible outcomes, $E_1$, $E_2$, …, $E_k$, so that the occurrence of an event $E_i$ is a Bernoulli random variable, $B_i$, with success probability $p_i$, where $\sum_{i=1}^{k} p_i = 1$. Assume all $p_i$ values are unknown, that $N$ trials of the experiment are conducted with $B_{ij} \coloneqq 1$ if the $j^{th}$ trial results in event $E_i$, and $B_{ij} \coloneqq 0$ otherwise, and define $f_i \coloneqq N - \sum_{j=1}^{N} B_{ij}$, the number of times event $E_i$ did not occur.
>
> A sufficient condition to be $1 - \alpha$ confident the most frequently observed outcome, $E_b$, is in fact the most likely, is that:

$$.5 < \frac{N - f_b + \frac{z^2}{2}}{N + z^2} - \frac{z}{N + z^2} \sqrt{\frac{f_b(N - f_b)}{N^3} + \frac{z^2}{4}}.$$

Formally, our algorithm proceeds as follows:

**Algorithm 1: Nested OCBA to Optimize the Decision Maker's Strategy**

1. Create an $|F_D|$-dimensional list, $B$, indicating the number of times each of the decision maker's strategies has been proposed by steps 2 through 18 as an optimum. Initiate all elements to 0.

    2. Create an $|F_D|$-dimensional list, $OCBA_D$, to store samples of $u_D(d, a, s)$ for all $d$.
    3. Initiate the number of samples for each $d$ in the initial iteration of OCBA to $N_d = N_{init}$, and set the iteration number to $iter_D = 0$.
    4. Increment $iter_D = iter_D + 1$.
    5. For all d:
        6. For all $i = 1:N_d$:
            7. Generate a sample, $r$, from $R$.

        **Determine the optimal adversarial strategy for $d, r$ using nested OCBA**
        8. Create an $|F_A|$-dimensional list, $OCBA_A$, to store samples of $U_A(d, a, s, r)$ for all $a$.
        9. Initiate the number of samples for each $a$ in the initial iteration of OCBA to $N_a = N_{init}$, and set the iteration number to $iter_A = 0$.
        10. Increment $iter_A = iter_A + 1$.
        11. For all $a$:
            12. For all $j = 1:N_a$:
                13. Generate a sample, $s_A$, from $S \sim Uni(h(d, a), h(d, a) + .1)$.
                14. Calculate $U_A(d, a, s_A, r)$ and store it in $OCBA_A(a)$.

        15. If $iter_A = 20$ or the stopping criterion of Appendix A3 is reached, set the adversary's optimal behavior for this value of $(d, r)$ to $a^*(d, r) :=$ the adversarial strategy with the highest sample mean in $OCBA_A$. Otherwise, use the OCBA allocation scheme, (1), to update $N_a$ for all $a$, and return to step 10. Note that when calculating (1), in the event the sample standard deviation is 0 for some strategy $i$ we set $s_i = .0001$. Values of $N$ are rounded up to get an integer number of samples, effectively setting $N_i = 1$.

        16. Generate a sample, $s_D$, from $S \sim Uni(h(d, a^*(d, r)), h(d, a^*(d, r)) + .1)$.
        17. Calculate $u_D(d, a^*(d, r), s_D)$ and store it in $OCBA_D(d)$.

18. If $iter_D = 20$ or the stopping criterion of Appendix A3 is reached, select the strategy $d$ with the highest sample mean in $OCBA_D$, and increment the element of $B$ corresponding to the selected optimum. Otherwise, use the OCBA allocation scheme, (1), to update $N_d$ for all $d$, and return to step 4 (see note in step 15 for the case when $s_i = 0$ for some strategy $i$).

19. If you are 95% confidence the most frequently observed strategy in $B$ is in fact the most likely outcome of a single trial (see lemma 1), stop; select the most frequently observed strategy in $B$ as the (approximated) optimal solution. Otherwise, repeat from step 2.

---

We present lemma 2 as a guide for estimating the expected number of trials for algorithm 1 to terminate, based on the required statistical confidence in algorithm 1's output:

> **Lemma 2 (see proof in Appendix A4)**
> Assume multiple trials of the experiment described in lemma 1 are to be performed and again define by $f_b$ the number of trials in which the most frequent outcome, $E_b$, does not occur. Given $f_b$, denote by $N(f_b)$ the minimum number of trials to be $1 - \alpha$ confident $E_b$ is in fact the most likely outcome of a single trial. Let $n_0 := \lceil z^2 \rceil$, where $z := \Phi^{-1}(1 - \alpha)$. We have the following:
>
> 1. Either $N(f_b) = 2f_b + n_0$, or $N(f_b) = 2f_b + n_0 + 1$.
> 2. A sufficient condition for $N(f_b) = 2f_b + n_0$ is: $\frac{\sqrt{3}}{18n_0} < \left(\frac{n_0}{2z}\right)^2 - \frac{z^2}{2}$.
> 3. Assuming the condition in part 2 is met and given $p_b$, the true probability of the most frequently observed outcome, the expected number of trials until algorithm 1 terminates can be calculated as:
>
> $$E[N(f_b)|p_b] = p_b^2 \cdot \sum_{f_b=0}^{\infty}(2f_b + n_0) \cdot P(!N(f_b - 1), f_b|p_b),$$
>
> where $P(!N(f_b - j), i|p_b) \cdot p_b^2$, the probability of not terminating after $N(f_b - j)$ trials while observing exactly $i$ failures during those trials, can be calculated recursively as:
>
> $P(!N(f_b - j), i|p_b) =$
> $\sum_{k=k_{min}}^{k_{max}} P(!N(f_b - j - 1), k|p_b) \cdot p_b^{2-i+k} \cdot (1 - p_b)^{i-k} \cdot \binom{2}{i - k},$
>
> where $k_{min} := \max\{f_b - j, i - 2\}$, $k_{max} = \min\{2(f_b - j - 1) + n_0, i\}$,
>
> and the recursion can be started using $P(!N(0), k|p_b) = \binom{n_0}{k} \cdot p_b^{n_0-k} \cdot (1 - p_b)^k$ for $1 \leq k \leq n_0$, and 0 otherwise.
>
> While this is an infinite sum it converges to its true value quickly as long as $p_b$ is sufficiently higher than .5.

The sufficient condition in part 2 of lemma 2 is met for the vast majority of confidence levels, including when using our required level of 95%. While the expected number of trials cannot be calculated without knowledge of $p_b$, a reasonable upper bound can be obtained by assuming any well-constructed algorithm will have $p_b \geq 60\%$; along with 95% required confidence, this gives $E[N(f_b)|.6] = 15$. In Appendix A5 we run many trials of our algorithm (more than required to terminate) and observe empirically that $p_b \approx 76.47\%$; this gives an expected number of trials of 5.7. As a technical note, to deal with situations where the estimated expected utilities of two or more strategies are essentially equivalent, in order to make the events $E_i$ distinct we assume the most frequently observed among them (in prior trials) is the optimum for the current trial. To avoid path dependencies we recalculate which is the most frequently observed after each trial. Ties are broken arbitrarily. The reason we require such a condition is that without it, two strategies that for all intents and purposes yield equivalent utilities will cause an excessive number of trials to be run in expectation, even once it's become clear the strategies are indistinguishable and either can be selected.

## 5. Discussion of findings

We analyzed 4 different cases: $n = 2, 3, 4,$ and 5, with corresponding parameters from Section 4 that are detailed in the next subsection. For each case we ran algorithm 1 to estimate the optimal solution to Colonel Blotto's problem, measured the time required to reach that solution, and performed ex post analysis using the sample utilities generated to calculate a lower bound on the probability our proposed optimum is in fact the true optimum. We also computed the exact optimal solution by exploiting the analytical forms of $\psi_D(d, a)$ and $\Psi_A(d, a)$ and the quadratic structure of $a^i(d) = \underset{a \in F_A}{\mathrm{argmax}}\, \Psi_A^i(d, a)$, and compared the results to the solution found by algorithm 1. For the remainder of this paper this solution method will simply be referred to as the "partial analytic" method. Even using the partial analytic method the computational advantages of algorithm 1 become apparent when $n \geq 4$. To make clear the computational savings when no analytical solutions are available (as will be the case in general) we also analyze the time required to solve the exact model when solving $\psi_D(d, a)$ and $\Psi_A(d, a)$ numerically and $a^i(d) = \underset{a \in F_A}{\mathrm{argmax}}\, \Psi_A^i(d, a)$ via full enumeration; this solution method will be referred to as the "fully numeric" method.

*5.i. Parameter values and sampling budgets*

As we increase $n$ from 2 to 5, we use the below parameter values. When $n = 2$, only the first two elements of each parameter are used, when $n = 3$ the first three elements are used, and so on.

$C_H = [.4,\ .35,\ .4,\ .4,\ .3]$, $c_A = [-.4984, -.4984, -.5529, -.6015, -.6834]$,
$c_D = [-.4984, -.4373, -.4373, -.5529, -.4626]$, $v = [1.3,\ .8,\ 1.25,\ .7,\ 1.1]$, and
$R = [R_1, R_2, R_3, R_4, R_5]$, where $R_1 \sim triangular(.8,\ 1,\ 1.5)$,
$R_2 \sim triangular(.5,\ .8,\ 2.5)$, $R_3 \sim triangular(1,\ 1.5,\ 3.5)$,
$R_4 \sim triangular(.3,\ .7,\ 1.1)$, and $R_5 \sim triangular(.6,\ 1.1,\ 1.9)$.

For algorithm 1 we set $N_{init} = 2^n$ as the number of samples to allocate to each of Colonel Blotto's strategies in the first iteration. In later iterations, a total of $5^n$ samples were allocated across the strategies per the OCBA allocation scheme. The same sampling rules were used for the nested OCBAs to solve Colonel Klink's decision problem.

*5.ii. Run times and ex post analysis*

We now compare the solve times using both the fully numeric method, the partial analytic method, and algorithm 1. Illustrating the difficulty in solving large ARA problems by brute force, the fully numeric method took 26 minutes when $n = 2$ and was computationally infeasible for $n \geq 3$. The partial analytic method took 1 minute when $n = 2$, 9 minutes when $n = 3$, 597 minutes (10 hours) when $n = 4$, and 62,000 minutes (43 days) when $n = 5$. In contrast, algorithm 1 took 1 minute when $n = 2$ and terminated after 3 trials, 24 minutes when $n = 3$ and terminated after 7 trials, 279 minutes (4.65 hours) when $n = 4$ and terminated after 5 trials, and 2,618 minutes (44 hours) when $n = 5$ and terminated after 3 trials. These results are summarized in table 1. The number or trials required for algorithm 1 to terminate is subject to random variation in the samples that are drawn, and in Appendix A5 we repeat the analysis many times for $n = 4$ to assess the expected number of trials to termination; in all cases algorithm 1 selects the correct optimum and the expected number of trials is rather low (5.7), but can be as high as 13.

Figure 2 shows graphically that our simulation-based algorithm identified the true optimum for all values of $n$. The true utilities of each strategy are plotted along with those estimated after the first iteration of the first trial of algorithm 1, and the estimated utilities upon termination of algorithm 1. Plotting the entire decision space for $n \geq 3$ would make the plots difficult to read so we've zeroed in on the true optimum; no estimated utilities of algorithm 1 (at termination) outside the range plotted exceeded the highest within the plotted range. The important thing to note about figure 2 is that while divergence between the true expected utilities and those computed by algorithm 1 remains in some parts of the decision space, algorithm 1 has converged near the peak values of expected utility, reflecting the computational budgetary allocation scheme that dedicates the majority of samples to understanding the most promising regions of the decision space. As can be seen in figures 2.b and 2.c, a false optimum would have been identified had the algorithm stopped after just 1 iteration of the first trial, but given additional iterations the algorithm arrived at the true optimum.

**Figure 2: exact values of $\psi_D(d)$ and estimates from algorithm 1**

— Exact ⋯•⋯ Simulation-based Solution, first iteration — - Simulation-based Solution, terminal

*Figure 2.a:  n = 2*

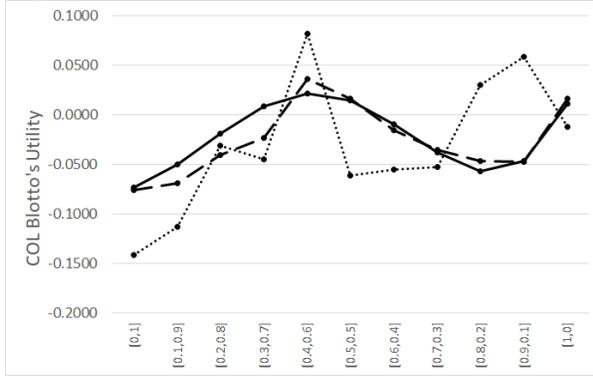

*Figure 2.b:  n = 3, near $d^* = [.7, 0, .3]$*

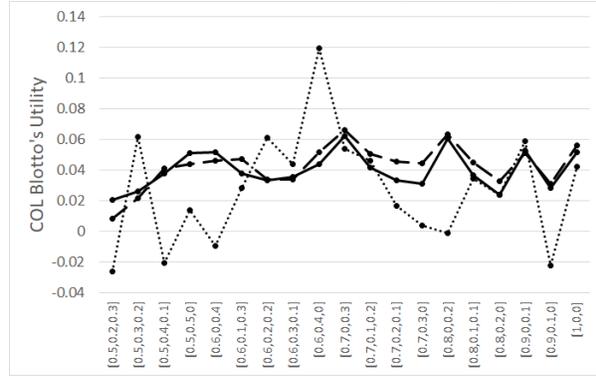

*Figure 2.c:  n = 4, near $d^* = [.4, 0, 0, .6]$*

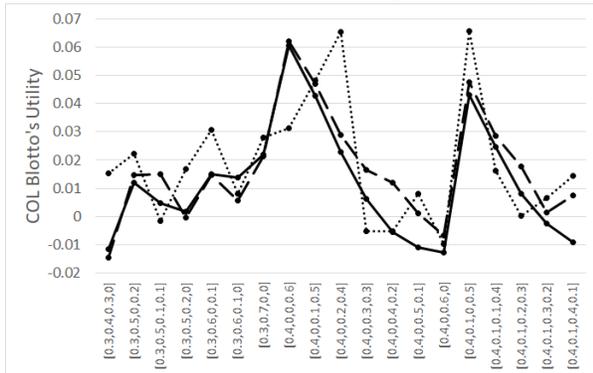

*Figure 2.d:  n = 5, near $d^* = [.4, 0, 0, .6, 0]$*

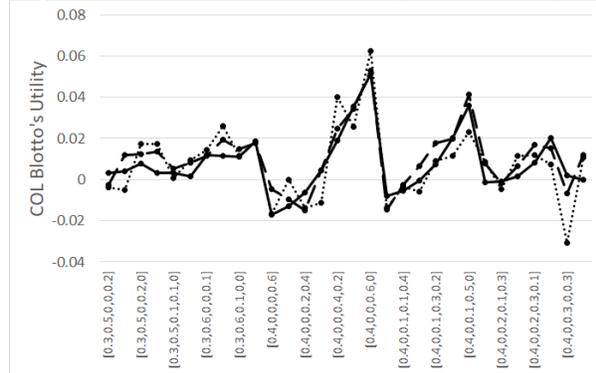

In general, we would not have an exact solution available to compare the results of our algorithm to, so to add confidence in our solution we used the sample values of utility, $u_D(d)$ for all $d \in F_D$, to calculate a lower bound on the probability the strategy with the highest sample mean utility, $d_b | \hat{u}_D(d_b) > \hat{u}_D(d_i) \ \forall \ i \neq b$, does in fact yield the highest utility. Using the assumption that $u_D(d)$ is normally distributed this is a straightforward calculation using the Bonferonni inequality described in equation (2). To justify the normal assumption we created normal qq-plots for $u_D(d)$ for a variety of values of $d$ and $n$. Figure 3 shows the plots for three particular values of $d$ when $n = 4$ which are illustrative of the results seen for other values; all showed clear normality despite the occasional bit of deviation in the tails (figure 3.c was the most extreme case found). Now, calculating the lower bounds on the probability the algorithm's proposed optimum is in fact the true optimum (APCS), we found that: in the case of $n = 2$, APCS evaluated to 91.41%; when $n = 3$, it was 57.01%; when $n = 4$ it's 94.59%; and when $n = 5$ it's 98.39%. These lower bounds are summarized in table 1. The markedly lower value of APCS when $n = 3$ is due to a point we noted in Section 4: as can be seen in figure 2.b, two strategies, $d_b = [.7, 0, .3]$ and $d_2 = [.8, 0, .2]$, have essentially identical expected utilities of .0660 and .0630, respectively. In the summation to calculate APCS, along with their respective standard errors (.0036 and .0049) we end up subtracting out $P(u_2 > u_b) = 31.12\%$. If we had not subtracted this term, essentially treating strategies $d_b$ and $d_2$ as one and the same, the lower bound on the probability of correct selection would have been 88.19%. In Appendix A5, when we repeat the analysis for $n = 4$ several times, we describe a simple calculation to adjust APCS values that doesn't require decision maker's to visually inspect expected utilities for virtual equivalency.

**Figure 3: normal qq-plot for $u_D(d)$ using 1$^{st}$ to 99$^{th}$ quantiles**

*Figure 3.a:* $d = [.4, 0, 0, .6]$     *Figure 3.b:* $d = [0, 0, .2, .8]$     *Figure 3.c:* $d = [.9, 0, .1, 0]$

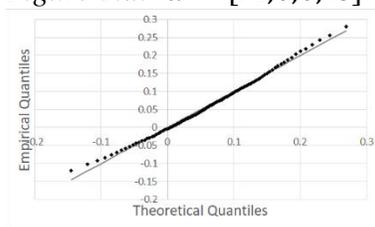 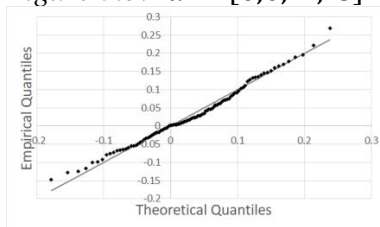 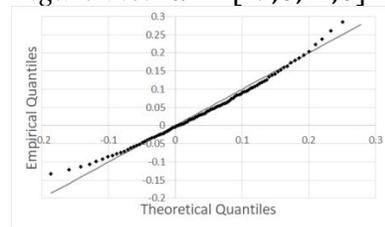

**Table 1: run times and lower bounds on probability of correct selection**

|  | n = 2 | n = 3 | n = 4 | n = 5 |
|---|---|---|---|---|
| Exact model solve time, without using analytical integrals over S and QIP | 26 | | | |
| Exact model solve time, using analytical integrals over S and QIP | 1 | 9 | 597 | 62,000 |
| Algorithm 1 solve time | 1 | 24 | 279 | 2,618 |
| Number of trials | 3 | 7 | 5 | 3 |
| Algorithm 1 PCS lower bound | 91.41% | 57.01% * | 94.59% | 98.39% |

All times measured in minutes
* When the two best strategies are considered equivalent, the bound on PCS is 88.19%

*5.iii. Benchmarking against a greedy search with random restarts*

As noted in Sections 1 and 2, there are few examples in the literature of solution techniques for large ARA, so the results presented in this paper are useful in their own right. However, we also compared our algorithm to that used by Banks, Rios, and Rios Insua (2016) to find a heuristic solution to a fairly large ARA model. They used a greedy algorithm with random restarts that begins with a randomly selected feasible strategy, generates a sufficient number of samples to estimate the expected utility of that strategy, and then permutes a single element of the strategy to reach a new feasible strategy. Samples are again drawn, and if this yields a new incumbent solution the algorithm continues with an additional permutation to this new incumbent; otherwise, the algorithm returns to the incumbent and tries an alternative permutation. Applying this to our Colonel Blotto example, we again have a nested greedy algorithm where for each strategy of Blotto he must draw random samples for Klink's $r$ value, and for each of these samples Blotto must use a greedy algorithm to analyze the decision problem from Klink's perspective. Using the case when $n = 4$, we assumed 100 samples was sufficient for each iteration of the algorithm; as observed empirically, a single trial of algorithm 1 generally allocates several hundred samples to each of the best few strategies, and 100 samples was seen to be a natural segregator between decent strategies and strategies that were entirely ignored following the initial iteration. As a stopping criterion we required the probability of finding a new local maximum with another random restart be less than 5% as calculated using Laplace's law of succession, which is consistent with algorithm 1's 95% required confidence level. This algorithm took 3,371 minutes (56 hours) to terminate with $n = 4$; as we detail in Appendix A5, in expectation our algorithm will take just 4.92 hours to terminate when $n = 4$. Further, though the greedy algorithm did find the true optimum, due to the relatively small sample size the APCS was -82.00% (i.e. no better than the trivial lower bound of 0%); while this is only a lower bound on the probability of correctly selecting the optimal solution, it clearly indicates much less confidence than when algorithm 1 is used. Indeed, as we show in Appendix A5 even after accounting for virtual equivalency in mean utilities, APCS for the greedy algorithm is less than 0%. We experimented with the number of samples to generate for each strategy in the greedy algorithm to determine how low the number needs to be in order for the algorithm to terminate in about 4.95 hours. We found the number of samples needed to be reduced to 7 per strategy. Of course, as APCS was meaningless when 100 samples per strategy were drawn, it's likewise meaningless with just 7 samples per strategy (-3,034.75%).

*5.iv. Summary of results*

Algorithm 1 accurately estimated $\psi_D(d)$ for all values of $n$ tested, and in Appendix A5 we repeat the analysis several times for $n = 4$ to assess the accuracy rate in selecting the true optimum. In brief, for the parameters in this section the algorithm always identifies the true optimum in 102 runs of the algorithm, and using three alternative parameter sets and an additional 144 runs, we find the algorithm always selects a strategy whose utility is at least 95% of the true optimum, though is typically much closer to 100% of optimality. Equally important to the accuracy of the algorithm is its speed, and the run times summarized in table 1 as well as Appendix A5 clearly show that our algorithm makes generically specified ARA problems that would otherwise be intractable solvable in a reasonable amount of time.

## 6. Conclusions and future work

This paper has presented a simulation-based algorithm for a two-player, sequential ARA model that terminates in reasonable time when the decision spaces for each the decision maker and adversary are large and the functional forms of utilities and state variable distributions do not offer analytical solutions. By running multiple independent trials, as defined by steps 2 through 18 of algorithm 1, we were able to terminate the algorithm only after sufficient statistical confidence in its output had been achieved. Further, in our example of Section 5 ex post analysis yielded lower bounds on the probability the algorithm produced the true optimum that were quite high. The example we used to assess our algorithm was a modified Colonel Blotto game and we used the number of battlefields as a variable parameter to increase the size of the model, but our methodology was not particular to Colonel Blotto games; the only criteria for our algorithm to be used is that the decision spaces are discrete and utilities can be reasonably approximated by a normal distribution.

What it means to solve a model quickly is subjective, and for our purposes we generally desired to be able to solve a model on a commercial laptop running Python in one day. Using this criterion we stopped analyzing our ARA model after $n = 5$, but with a more powerful machine and/or willingness to wait for solutions, larger problems could be solved. The main result of this paper is that our OCBA-based algorithm offers enormous computational savings for ARA models and presents a framework for exploring the computational feasibility of increasingly larger models. A natural next step is to develop methods to identify promising regions of very large decision spaces, within which algorithm 1 can be applied. Three general approaches come to mind for exploring this problem: assess all possible decisions using a low-fidelity model, as in Xu, et al. (2016), and then explore in detail regions where the low-fidelity model suggested that expected utility was high; partition the decision space and draw samples of decisions from each, which are then explored in detail to decide whether the region as a whole ought to be explored, as in Chen, et al. (2014); and calibrating a distribution for the expected utility of a pair of decision maker and adversary decisions, $g(d, a, s) = u_D(d, a, s) \cdot p(s|d, a)$, via a Markov Chain Monte Carlo simulation, as in Bielza, et al. (1999), and then analyzing the marginal of the distribution with respect to $d$ to determine regions where the expected utility is high. All of these approaches will have to deal with the unique challenge of ARA models, which is that the adversary's utility function and beliefs are stochastic. In this paper we overcame this challenge by utilizing a nested OCBA algorithm.

Another natural extension of our work would be to incorporate a richer description of uncertainty regarding the adversary's behavior. Considering that we estimated the distribution of the adversary's behavior, $P_D(a|d)$, by solving his problem using statistical selection, despite the accuracy of our algorithm seen herein it is still subject to sampling error. By using simple point estimates for each $P_D(a_i|d)$, $a_i \in F_A$, sampling error could lead us to assume the adversary is using a suboptimal strategy, which in turn leads us to overestimate the decision maker's utility from each strategy and may even suggest a false optimum. While it's reasonable to base decisions around our algorithm because, by theorem 1, when analyzing the adversary's problem it maximizes the probability of predicting his true optimum, risk-averse decision makers may instead wish to characterize the uncertainty in each $P_D(a_i|d)$ and employ a maximin strategy that maximizes utility subject to the worst (reasonable) case for the values of $P_D(a_i|d)$, even at the expense of lower expected utility. Such an approach might involve estimating confidence intervals for each $P_D(a_i|d)$ and then using these as uncertainty sets in a distributionally robust optimization. Distributional robustness in ARA models was studied in McLay, et al. (2012), but in their analysis uncertainty sets were taken as given rather than estimated analytically. Intuitively, heuristically defined uncertainty sets such as $P_D(a_i|d) \in \{\hat{P}_D(a_i|d) \pm 5\%\}$, where $\hat{P}_D(a_i|d)$ are the point estimates used in this paper, could be employed, but ideally future research will formalize rigorous methodologies to determine uncertainty sets.

Lastly, while algorithm 1 was seen to outperform an alternative which is the only algorithm to our knowledge that's been employed to a decently sized ARA problem, further alternatives should continue to be explored as computational ARA is a very young field. For example, the alternative classes of statistical selection techniques mentioned in Section 2 could be employed in the ARA context and would serve as natural benchmarks for our results. This will allow our algorithm to be compared to other sophisticated methodologies to gain a better understanding of its value in solving this young class of game theoretic models.

Lapan, Harvey E., and Todd Sandler (1988) To Bargain or Not To Bargain: That Is The Question. *The American Economic Review* 78(2):16–21.

Liang, Xiannuan, and Yang Xiao (2013) Game Theory for Network Security. *IEEE Communications Surveys & Tutorials* 15(1):472–86.

Lin, Kyle, and Jeffrey Dayton (2011) Game-Theoretic Models for Jamming Radio-Controlled Improvised Explosive Devices. *Military Operations Research* 16(3):5–13.

McLay, Laura, Casey Rothschild, and Seth Guikema (2012) Robust Adversarial Risk Analysis: A Level-k Approach. *Decision Analysis* 9(1):41–54.

Neumann, John von (1928) Zur Theorie Der Gesellschaftsspiele. *Mathematical Annals* 100.

Nikoofal, Mohammad E., and Jun Zhuang (2012) Robust Allocation of a Defensive Budget Considering an Attacker's Private Information: Robust Allocation of a Defensive Budget. *Risk Analysis* 32(5):930–43.

Peng, Yijie, Chun-Hung Chen, Michael C. Fu, and Jian-Qiang Hu (2016) Dynamic Sampling Allocation and Design Selection. *INFORMS Journal on Computing* 28(2):195–208.

Rios Insua, David, Jesus Rios, and David Banks (2009) Adversarial Risk Analysis. *Journal of the American Statistical Association* 104(486):841–54.

Rios, Jesus, and David Rios Insua (2012) Adversarial Risk Analysis for Counterterrorism Modeling. *Risk Analysis* 32(5):894–915.

Sevillano, Juan Carlos, David Rios Insua, and Jesus Rios (2012) Adversarial Risk Analysis: The Somali Pirates Case. *Decision Analysis* 9(2):86–95.

Wang, Chen, and Vicki M. Bier (2011) Target-Hardening Decisions Based on Uncertain Multiattribute Terrorist Utility. *Decision Analysis* 8(4):286–302.

Wang, Shouqiang, and David Banks (2011) Network Routing for Insurgency: An Adversarial Risk Analysis Framework. *Naval Research Logistics (NRL)* 58(6):595–607.

Wilson, Edwin B. (1927) Probable Inference, the Law of Succession, and Statistical Inference. *Journal of the American Statistical Association* 22(158):209–12.

Xu, Jie, Si Zhang, Edward Huang, Chun-Hung Chen, Loo Hay Lee, and Nurcin Celik (2016) Multi-fidelity Optimization with Ordinal Transformation and Optimal Sampling. *Asia-Pacific Journal of Operational Research* 33(3):26.

Zhuang, Jun, and Vicki M. Bier (2007) Balancing Terrorism and Natural Disasters—Defensive Strategy with Endogenous Attacker Effort. *Operations Research* 55(5):976–91.
33

# Appendix

## A1. Closed form solutions for $\psi_D(d, a)$ and $\Psi_A(d, a)$

We show here that $\psi_D(d, a)$ and $\Psi_A(d, a)$ can be solved analytically using the distributions and utility functions specified in Section 4.i.

$$\psi_A(d, a) = \int_S U_A(d, a, s, r) p(s|d, a) ds$$

$$= \int_{h_n}^{h_n+.1} \int_{h_{n-1}}^{h_{n-1}+.1} \cdots \int_{h_2}^{h_2+.1} \int_{h_1}^{h_1+.1} \frac{\sum_{i=1}^{n} r_i\left(1 - e^{-4.6(s_i - C_{H,i} - .05)}\right)}{n} \cdot 10^n \, ds_1 \ldots ds_n$$

(since $S_i | d, a \sim Uni(h_i, h_i + .1)$ for all $i$)

$$= \int_{h_n}^{h_n+.1} \cdots \int_{h_2}^{h_2+.1} \left[ \int_{h_1}^{h_1+.1} \frac{r_1\left(1 - e^{-4.6(s_1 - C_{H,1} - .05)}\right)}{n} \cdot 10 \, ds_1 \right] \cdot$$

$$\frac{\sum_{i \neq 1} r_i\left(1 - e^{-4.6(s_i - C_{H,i} - .05)}\right)}{n} \cdot 10^{n-1} ds_2 \ldots ds_n.$$

Focusing just on the term inside the [ ], we obtain:

$$\int_{h_1}^{h_1+.1} \frac{r_1\left(1 - e^{-4.6(s_1 - C_{H,1} - .05)}\right)}{n} \cdot 10 \, ds_1 = \frac{r_1}{n} \cdot \left(1 - 10 \cdot e^{4.6(C_{H,1} + .05)} \left[-\frac{1}{4.6} e^{-4.6 s_1}\right]_{h_1}^{h_1+.1}\right)$$

$$= \frac{r_1}{n} \cdot \left(1 + \frac{10}{4.6} \cdot e^{4.6(C_{H,1} + .05)} \cdot \left(e^{-4.6(h_1 + .1)} - e^{-4.6 h_1}\right)\right)$$

$$= \frac{r_1}{n} \cdot \left(1 + \frac{10}{4.6} \cdot e^{4.6(C_{H,1} + .05)} \cdot \left(e^{-.46} - 1\right) \cdot e^{-4.6 h_1}\right)$$

$$:= I_1^A. \tag{3}$$

From here, it's easy to show that repeatedly integrating over the elements of $S$ yields:

$$\psi_A(d, a) = \sum_{i=1}^{n} I_i^A.$$

At the time $\psi_A(d, a)$ must be evaluated, $d$ and $a$ are assumed known and hence $h_i$ is known for all $i$. $r_i$ is also assumed known at this point, and everything $C_{H,i}$ is a given constant. Hence, $\sum_{i=1}^{n} I_i^A$ can be calculated by direct computation. The defender's utility function has an almost identical form as the attacker's, and it's easy to show that $\psi_D(d, a) = \sum_{i=1}^{n} I_i^D$, where:

$$I_i^D := -\frac{v_i}{n} \cdot \left(1 + \frac{10}{4.6} \cdot e^{4.6(C_{H,1} + .05)} \cdot \left(e^{-.46} - 1\right) \cdot e^{-4.6 h_1}\right).$$

## A2. Formulation of $a^*(d) = argmax_a \psi_A(d, a)$ as a quadratic integer program

At node A of figure 1.b, the attacker's problem is to solve:

$$\max_a z = \sum_{i=1}^n I_i^A \qquad (4)$$
$$s.t.$$
$$\sum_{i=1}^n a_i = 1$$
$$a \in \{0, .1, .2, \ldots, .9, 1\}.$$

Simply plugging in the formula for $h_i$ given in Section 4.i into (3) yields:

$$I_i^A = \frac{r_i}{n} \cdot \left(1 + \frac{10}{4.6} \cdot e^{4.6(C_{H,i}+.05)} \cdot (e^{-.46} - 1) \cdot e^{-4.6h_i}\right)$$
$$= A_{1,i} + A_{2,i} \cdot e^{-4.6h_i}$$
$$\text{where } A_{1,i} := \frac{r_i}{n} \text{ and } A_{2,i} := A_{1,i} \cdot \frac{10}{4.6} \cdot e^{4.6(C_{H,i}+.05)} \cdot (e^{-.46} - 1).$$
$$= A_{1,i} + A_{2,i} \cdot e^{-4.6 \cdot -\frac{2}{4.6} \cdot \log\left(\frac{c_A^i a_i + 1}{c_D^i d_i + 1}\right) - 4.6 C_H^i}$$
$$= A_{1,i} + A_{2,i} \cdot e^{-4.6 C_H^i} \cdot \left[e^{\log\left(\frac{c_A^i a_i + 1}{c_D^i d_i + 1}\right)}\right]^2$$
$$= A_{1,i} + A_{2,i} \cdot e^{-4.6 C_H^i} \cdot (c_D^i d_i + 1)^{-2} (c_A^i a_i + 1)^2,$$

which is quadratic in $a_i$.

Therefore, model (4) is a quadratic program with a discrete feasible region. Making the transformation $a_i \to 10 a_i$ and adjusting the objective function accordingly gives a small quadratic integer program which can be solved using off-the-shelf solvers.

## A3. Stopping criterion for each trial of algorithm 1

In steps 15 and 18 of algorithm 1, we use the following stopping criteria based on a weighted sum of percentage changes in strategies across iterations. The weights applied to each strategy are the number of samples allocated in the most recent iteration, normalized to sum to 1: $w_i = N_i / \sum_i N_i$. This places the vast majority of the weight only on those strategies we reasonably think may be the optimum. Using these weights we calculate a weighted average percentage change between the current and each of the last 4 iterations, where percentage changes are calculated with respect to the difference between the best and the worst expected utility at the current iteration, $\delta := max_i \alpha_i - min_i \alpha_i$, where as in Section 4.iii, $\alpha_i$ represents the sample expected utility for strategy $i$. This choice was made because dividing by expected utility itself causes unduly high percentage changes when $\alpha_i$ is near 0. Adding a subscript for the iteration number and assuming the current iteration is $t$, we calculate the weighted percentage change between the current iteration and iteration $t'$ as: $q_{t,t'} := \sum_i w_{t,i} \cdot |\alpha_{t,i} - \alpha_{t',i}| / \delta_t$. Our stopping criteria is that $q_{t,t'} \leq .05$ for $t' = t-1, t-2, t-3$, and $t-4$, indicating no significant changes over the last 4 iterations.

### A4. Proof of lemmas 1 and 2

*A4.i. Lemma 1*

Wilson (1927) gives $1 - \alpha$ one-sided confidence intervals for $p_i$ of:

$$p_i \geq \frac{N - f_i + \frac{z^2}{2}}{N + z^2} - \frac{z}{N + z^2} \sqrt{\frac{f_i(N - f_i)}{N^3} + \frac{z^2}{4}}, \text{ where } z := \Phi^{-1}(1 - \alpha).$$

By definition if $p_b > .5$, then $E_b$ is the most likely outcome, and therefore if we're $1 - \alpha$ confident $p_b > .5$ then we're at least $1 - \alpha$ confident $E_b$ is the most likely outcome. We say we're "at least" $1 - \alpha$ confident, and that this is a sufficient but not necessary condition, because it could be that $p_i < .5$ for all $i$.

## A4.ii. Lemma 2

We first prove part 1 of the lemma by showing that: (1a) $2f_b + n_0 - 1 < N(f_b)$; and (1b) $N(f_b) \leq 2f_b + n_0 + 1$. Using Wilson's confidence intervals, part 1a is equivalent to showing:

$$\frac{f_b+n_0-1+z^2/2}{2f_b+n_0-1+z^2} - \frac{z}{2f_b+n_0-1+z^2}\sqrt{\frac{f_b(f_b+n_0-1)}{(2f_b+n_0-1)^3} + \frac{z^2}{4}} < .5 \leftrightarrow$$

$$\frac{n_0-1}{2} - z\sqrt{\frac{f_b(f_b+n_0-1)}{(2f_b+n_0-1)^3} + \frac{z^2}{4}} < 0 \leftrightarrow$$

$$\frac{f_b(f_b+n_0-1)}{(2f_b+n_0-1)^3} > \left(\frac{n_0-1}{2z}\right)^2 - \frac{z^2}{4}$$

We define $g_c(x) = \frac{x(x+c)}{(2x+c)^3}$, which will be used again when proving parts 1b and 2 of the lemma.

If $x, c \geq 0$, then $g(x) \geq 0$, and since $f_b, n_0 - 1 \geq 0$, if we show that $0 > \left(\frac{n_0-1}{2z}\right)^2 - \frac{z^2}{4}$ we'll have proved part 1a. Note that:

$$\left(\frac{n_0-1}{2z}\right)^2 - \frac{z^2}{4} = \frac{1}{4}\left(\frac{n_0-z^2-1}{z}\right)\left(\frac{n_0+z^2-1}{z}\right)$$

Clearly, $\frac{n_0+z^2-1}{z} > 0$. Further, $n_0 - z^2 < 1$ and therefore $\frac{n_0-z^2-1}{z} < 0$. It follows that

$$\frac{1}{4}\left(\frac{n_0-z^2-1}{z}\right)\left(\frac{n_0+z^2-1}{z}\right) = \left(\frac{n_0-1}{2z}\right)^2 - \frac{z^2}{4} < 0,$$ and therefore part 1a has been proved.

Similarly to part 1a, part 1b is true if $g_{n_0+1}(f_b) \leq \left(\frac{n_0+1}{2z}\right)^2 - \frac{z^2}{4}$ for all $f_b$. For any value of $c > 0$, calculus shows that $g_c(x)$ attains its maximum value at $x = \frac{c}{2}(\sqrt{3} - 1)$, and $g_c\left(\frac{c}{2}(\sqrt{3} - 1)\right) = \frac{\sqrt{3}}{18c}$. Therefore, to prove part 1b it's sufficient to show $\frac{\sqrt{3}}{18(n_0+1)} \leq \left(\frac{n_0+1}{2z}\right)^2 - \frac{z^2}{4}$. Observe:

$$\frac{\sqrt{3}}{18(n_0+1)} \leq \left(\frac{n_0+1}{2z}\right)^2 - \frac{z^2}{4} \leftrightarrow$$

$$\frac{2\sqrt{3}}{9} \leq (n_0 + 1)\left(\frac{n_0+1}{z} - z\right)\left(\frac{n_0+1}{z} + z\right) \leftrightarrow$$

$$\frac{2\sqrt{3}}{9} \leq (n_0 + 1)\left(\frac{n_0-z^2+1}{z}\right)\left(\frac{n_0+z^2+1}{z}\right) \leftrightarrow$$

$$\frac{2\sqrt{3}}{9} \leq \left(\frac{n_0+1}{z}\right)(n_0 - z^2 + 1)\left(\frac{n_0+z^2+1}{z}\right) \tag{5}$$

Noting that each term in the product on the righthand side of (5) is greater than 1 and that $\frac{2\sqrt{3}}{9} = .3849 < 1$, it follows that (5) is true, and therefore 1b is true.

Part 2 of the lemma is a natural extension of part 1. We already know $N(f_b) > 2f_b + n_0 - 1$ by part 1a. To show $N(f_b) = 2f_b + n_0$, following the same reasoning as part 1b a sufficient condition is that $\frac{\sqrt{3}}{18n_0} \leq \left(\frac{n_0}{2z}\right)^2 - \frac{z^2}{4}$. The reason this is not a necessary condition is that the point where $g_{n_0}(f_b)$ attains its maximum, $f_{b,max} = \frac{n_0}{2}(\sqrt{3} - 1)$, may not be an integer, and the number of failures by definition is an integer. A necessary and sufficient condition for $N(f_b) = 2f_b + n_0$ is that $\max\{g_{n_0}(f_{b-}), g_{n_0}(f_{b+})\} \leq \left(\frac{n_0}{2z}\right)^2 - \frac{z^2}{4}$, where $f_{b-} := \left\lfloor\frac{\sqrt{3}}{18n_0}\right\rfloor$ and $f_{b+} := \left\lceil\frac{\sqrt{3}}{18n_0}\right\rceil$; this is true because $g_c(x)$ is unimodal for $x, c \geq 0$.

To prove part 3 we must show that: (3a) $P(N(f_b)|p_b) = P(!N(f_b - 1), f_b|p_b) \cdot p_b^2$; and (3b) $P(!N(f_b - j), i|p_b) = \sum_{k=k_{min}}^{k_{max}} P(!N(f_b - j - 1), k|p_b) \cdot p_b^{2-i+k} \cdot (1 - p_b)^{i-k} \cdot \binom{2}{i-k}$. To see that 3a is true, note that $N(f_b) - N(f_b - 1) = 2f_b + n_0 - (2(f_b - 1) + n_0) = 2$, and therefore if $f_b$ failures have already been observed through the first $N(f_b - 1)$ trials the next two must both be successes, and hence $P(!N(f_b - 1), f_b|p_b)$ is multiplied by $p_b^2$. If anything less than $f_b$ failures had been observed through the first $N(f_b - 1)$ trials, then the algorithm would have terminated because, by assumption, only $N(f_b - 1)$ trials are required to terminate the algorithm when $f_b - 1$ failures have been observed, and thus the equality in 3a holds. To prove part 3b, first note that, given $i$ failures occurred through the first $N(f_b - j)$ trials, the minimum number of failures through $N(f_b - j - 1)$ is the larger of $i - 2$ and the least number of failures that prevents the algorithm from terminating after $f_b - j - 1$ trials, or $f_b - j$. Hence we've defined $k_{min} = \max\{f_b - j, i - 2\}$. Also note the maximum failures through $N(f_b - j - 1)$ trials is the smaller of $N(f_b - j - 1) = 2(f_b - j - 1) + n_0$ and the number of failures we're given has occurred through $f_b - j$ trials, or $i$, and hence $k_{max} := \min\{2(f_b - j - 1) + n_0, i\}$. Now 3b holds as a matter of definition, since there are 2 trials between $N(f_b - j - 1) + 1$ and $N(f_b - j)$, inclusive, the number of failures occurring between these trials is $i - k$, and there are $\binom{2}{i - k}$ ways for $i - k$ failures to occur.

## A5. Robustness of algorithm 1's results

In Section 5 we saw that for each value of $n$, a single run of our algorithm produced the true optimum in each case. Also recall that when $n = 2$ the algorithm terminated after 3 trials (and hence no trials produced a false optimum), when $n = 3$, 7 trials were required (and hence 2 failures occurred), when $n = 4$, 5 trials were needed (hence, 1 failure), and when $n = 5$ the algorithm required 3 trials (0 failures). To gain further understanding as to our algorithm's ability to produce the true optimum and the required number of trials to terminate, we reran algorithm 1 several times for the case when $n = 4$. Together with the results from Section 5 we ran the algorithm a total of 18 times for $n = 4$ using the model parameters in Section 5, which amounted to 102 total trials. In all 18 runs of the algorithm the true optimum was identified, but the number of trials required for termination could vary significantly. The required number was concentrated towards the low end, with a mode of just 3 and an average of 5.7, but the maximum required was 13, which occurred only once. Empirically, over 102 trials the true optimum was selected 76.47% of the time. Also of interest is the average time required for each trial; this was 52 minutes, and hence in Section 5.iii we benchmarked our algorithm against a greedy search that took approximately $5.7 \cdot 52/60 = 4.92$ hours to terminate.

To add further credence to our algorithm we also reran the algorithm 10 times for three alternative parameter sets when $n = 4$, listed below. The alternative parameter sets are listed below, and the required number of trials and time to termination are given in table 2. Table 2 also shows the true utility of the strategy selected by algorithm 1, given as a percentage of the true optimum; the majority of the time the algorithm selects the true optimum, but when it does not the true utility of the strategy selected is at a minimum 95.60% of the true optimum. Lastly, using only the samples generated (i.e. *not* using knowledge of the true optimum) table 2 lists the lower bounds on the probability of correctly selecting the optimal strategy (APCS). As we noted in Section 5.ii, APCS can be a poor metric when multiple strategies provide nearly equivalent utilities. To provide decision makers with a sensible measurement that assumes ambivalence between strategies that are virtually equivalent, we define the following modified version of equation (2):

$$APCS(x) := 1 - \sum_{i \in I} \Phi\left(\frac{\hat{u}_D(d_i) - \hat{u}_D(d_i)}{\sqrt{se_b^2 + se_i^2}}\right), \qquad (6)$$

where $se_i$ is the standard error of strategy $i$, $I := \left\{i \mid \frac{\hat{u}_D(d_i) - \hat{u}_D(d_w)}{\hat{u}_D(d_b) - \hat{u}_D(d_w)} < x\right\}$, $d_w$ is the strategy with the lowest sample expected utility, and as before $d_b$ is that with the highest sample expected utility.

Equation (6) effectively considers any strategy that's within $100 \cdot x$ percent of the highest sample expected utility to be an acceptable strategy choice. We computed $APCS(x)$ for $x \in \{99\%, 98\%, 97\%, 96\%, 95\%\}$ and report the results in table 2.

> *Original parameters*
> $C_H = [.4, .35, .4, .4]$, $c_A = [-.4984, -.4984, -.5529, -.6015]$,
> $c_D = [-.4984, -.4373, -.4373, -.5529]$, $v = [1.3, .8, 1.25, .7]$, and
> $R = [R_1, R_2, R_3, R_4]$, where $R_1 \sim triangular(.8, 1, 1.5)$, $R_2 \sim triangular(.5, .8, 2.5)$, $R_3 \sim triangular(1, 1.5, 3.5)$, and $R_4 \sim triangular(.3, .7, 1.1)$.
>
> *Parameter set 2: mirroring scenario. In this scenario, the modes in the triangle distributions for R equal v, and the upper and lower bounds are $\pm.3$ from the modes.*
> $C_H = [.4, .35, .4, .4]$, $c_A = [-.4984, -.4984, -.5529, -.6015]$,
> $c_D = [-.4982, -.4373, -.4373, -.5529]$, $v = [1.3, .8, 1.25, .7]$, and
> $R = [R_1, R_2, R_3, R_4]$, where $R_1 \sim triangular(1, 1.3, 1.6)$, $R_2 \sim triangular(.5, .8, 1.1)$, $R_3 \sim triangular(.95, 1.25, 1.55)$, and $R_4 \sim triangular(.4, .7, 1)$.
>
> *Parameter set 3: low incentive scenario. The status quo ($C_H$) is relatively good for the decision maker at all but her least desirable battlefield (battlefield 4).*
> $C_H = [.33, .33, .33, .4]$, $c_A = [-.5730, -.5210, -.6194, -.6015]$,
> $c_D = [-.4108, -.4108, -.3390, -.5529]$, $v = [1.3, .8, 1.25, .7]$, and
> $R = [R_1, R_2, R_3, R_4]$, where $R_1 \sim triangular(.8, 1, 1.5)$, $R_2 \sim triangular(.5, .8, 2.5)$, $R_3 \sim triangular(1, 1.5, 3.5)$, and $R_4 \sim triangular(.3, .7, 1.1)$.
>
> *Parameter set 4: randomized scenario. These parameters bear no relation to the original parameter set.*
> $C_H = [.39, .35, .37, .34]$, $c_A = [-.5827, -.5210, -.5529, -.5730]$,
> $c_D = [-.5631, -.3540, -.4242, -.4982]$, $v = [2.3, 1.7, 3.4, 2.1]$, and
> $R = [R_1, R_2, R_3, R_4]$, where $R_1 \sim triangular(2.8, 3, 3.1)$, $R_2 \sim triangular(2.5, 3.4, 3.5)$, $R_3 \sim triangular(.8, 1.1, 1.9)$, and $R_4 \sim triangular(2.9, 3.2, 3.5)$.

**Table 2: required trials, times, and lower bounds on probability of correct selection**

| Parameter Set | Run | Trials | Time (hours) | % Optimum | APCS | APCS(99%) | APCS(98%) | APCS(97%) | APCS(96%) | APCS(95%) |
|---|---|---|---|---|---|---|---|---|---|---|
| Original | 1 | 5 | 4.6209 | 100.00% | 94.5781% | 94.5781% | 94.5781% | 94.5781% | 98.8322% | 98.8322% |
| Original | 2 | 3 | 2.3290 | 100.00% | 86.4522% | 86.4522% | 86.4522% | 86.4522% | 86.4522% | 88.6974% |
| Original | 3 | 5 | 4.1744 | 100.00% | 88.9501% | 88.9501% | 96.6762% | 96.6762% | 96.6762% | 96.6762% |
| Original | 4 | 3 | 2.5943 | 100.00% | 65.9286% | 65.9286% | 65.9286% | 65.9286% | 67.3257% | 72.0249% |
| Original | 5 | 3 | 3.1451 | 100.00% | 91.9086% | 91.9086% | 91.9086% | 96.3086% | 96.3086% | 96.3086% |
| Original | 6 | 3 | 2.6621 | 100.00% | 48.1133% | 88.0593% | 88.0593% | 88.0593% | 88.0593% | 88.0593% |
| Original | 7 | 9 | 7.4437 | 100.00% | 98.6345% | 98.6345% | 98.6345% | 98.6345% | 99.9408% | 99.9408% |
| Original | 8 | 5 | 4.3114 | 100.00% | 98.3496% | 98.3496% | 98.3496% | 98.3496% | 98.3496% | 99.8269% |
| Original | 9 | 7 | 6.0717 | 100.00% | 82.9616% | 82.9616% | 82.9616% | 99.8365% | 99.8365% | 99.8365% |
| Original | 10 | 5 | 4.0994 | 100.00% | 93.3289% | 93.3289% | 93.3289% | 93.3289% | 94.1222% | 98.8892% |
| Original | 11 | 3 | 2.7186 | 100.00% | 51.7583% | 51.7583% | 51.7583% | 70.9374% | 78.3933% | 80.5176% |
| Original | 12 | 11 | 9.5211 | 100.00% | 96.5826% | 96.5826% | 99.6875% | 99.9922% | 99.9922% | 99.9922% |
| Original | 13 | 3 | 2.4231 | 100.00% | 72.9906% | 72.9906% | 87.7637% | 87.7637% | 87.7637% | 87.7637% |
| Original | 14 | 5 | 4.7192 | 100.00% | 94.6204% | 94.6204% | 94.6204% | 94.6204% | 97.3383% | 97.4005% |
| Original | 15 | 13 | 11.0425 | 100.00% | 98.1009% | 98.1009% | 98.1009% | 99.8232% | 99.9682% | 99.9977% |
| Original | 16 | 9 | 7.6797 | 100.00% | 99.0674% | 99.0674% | 99.0674% | 99.9343% | 99.9343% | 99.9590% |
| Original | 17 | 7 | 6.2415 | 100.00% | 98.5667% | 98.5667% | 98.5667% | 98.5667% | 99.8249% | 99.8844% |
| Original | 18 | 3 | 2.6933 | 100.00% | 58.1801% | 58.1801% | 58.1801% | 70.5259% | 70.5259% | 72.6036% |
| 2 | 1 | 3 | 2.4675 | 100.00% | 99.9025% | 99.9025% | 99.9025% | 99.9025% | 99.9025% | 99.9025% |
| 2 | 2 | 3 | 2.5464 | 100.00% | 99.9719% | 99.9719% | 99.9719% | 99.9719% | 99.9719% | 99.9719% |
| 2 | 3 | 3 | 2.7767 | 100.00% | 99.9982% | 99.9982% | 99.9982% | 99.9982% | 99.9982% | 99.9982% |
| 2 | 4 | 3 | 2.5408 | 100.00% | 99.9464% | 99.9464% | 99.9464% | 99.9464% | 99.9464% | 99.9464% |
| 2 | 5 | 3 | 2.6114 | 100.00% | 99.9463% | 99.9463% | 99.9463% | 99.9463% | 99.9463% | 99.9463% |
| 2 | 6 | 3 | 2.8472 | 100.00% | 99.9823% | 99.9823% | 99.9823% | 99.9823% | 99.9823% | 99.9823% |
| 2 | 7 | 3 | 2.7889 | 100.00% | 99.9977% | 99.9977% | 99.9977% | 99.9977% | 99.9977% | 99.9977% |
| 2 | 8 | 5 | 4.1136 | 100.00% | 99.2218% | 99.2218% | 99.2218% | 99.2218% | 99.2218% | 99.9732% |
| 2 | 9 | 3 | 2.4742 | 100.00% | 99.9795% | 99.9795% | 99.9795% | 99.9795% | 99.9795% | 99.9795% |
| 2 | 10 | 3 | 2.4822 | 100.00% | 99.9357% | 99.9357% | 99.9357% | 99.9357% | 99.9357% | 99.9357% |
| 3 | 1 | 5 | 4.2419 | 98.85% | 20.7777% | 62.9565% | 97.8979% | 97.8979% | 97.8979% | 97.8979% |
| 3 | 2 | 3 | 2.5672 | 99.82% | 47.6974% | 47.6974% | 69.1484% | 79.8305% | 81.9716% | 94.8479% |
| 3 | 3 | 3 | 2.3906 | 99.75% | 93.2065% | 93.2065% | 93.2065% | 96.2494% | 96.9947% | 99.1739% |
| 3 | 4 | 11 | 8.8308 | 99.75% | 74.1395% | 74.1395% | 98.9298% | 99.7562% | 99.9864% | 99.9995% |
| 3 | 5 | 7 | 5.9297 | 99.75% | 20.7633% | 99.4819% | 99.4819% | 99.4819% | 99.9517% | 99.9517% |
| 3 | 6 | 3 | 2.5475 | 100.00% | -2.3521% | 65.6943% | 86.5776% | 86.5776% | 90.7887% | 96.7025% |
| 3 | 7 | 5 | 4.1925 | 100.00% | 73.1479% | 73.1479% | 94.4568% | 98.8500% | 98.8500% | 99.3940% |
| 3 | 8 | 3 | 2.4447 | 100.00% | 28.7584% | 66.4094% | 92.8074% | 92.8074% | 93.8089% | 95.2626% |
| 3 | 9 | 3 | 2.3878 | 100.00% | 35.7291% | 35.7291% | 74.9243% | 86.2074% | 86.2074% | 91.3297% |
| 3 | 10 | 3 | 2.6775 | 100.00% | -10.8486% | 68.6894% | 88.7759% | 88.7759% | 94.3021% | 94.3021% |
| 4 | 1 | 5 | 4.2408 | 99.84% | 5.9631% | 68.2740% | 76.6379% | 91.0776% | 98.2393% | 99.9059% |
| 4 | 2 | 3 | 2.3983 | 95.60% | 36.1192% | 36.1192% | 58.7280% | 72.0863% | 90.5764% | 96.8366% |
| 4 | 3 | 3 | 2.3967 | 99.84% | -49.0188% | 62.2326% | 62.2326% | 72.3289% | 88.1562% | 88.1562% |
| 4 | 4 | 3 | 2.8375 | 100.00% | -64.3456% | 27.3455% | 66.3419% | 81.2045% | 91.5331% | 91.5331% |
| 4 | 5 | 3 | 2.4619 | 100.00% | -65.9713% | 14.3660% | 14.3660% | 54.6571% | 93.9504% | 93.9504% |
| 4 | 6 | 7 | 6.5617 | 96.72% | 48.0206% | 74.3938% | 94.3866% | 99.5608% | 99.7396% | 99.7656% |
| 4 | 7 | 9 | 7.9350 | 100.00% | 16.8710% | 73.2499% | 98.9508% | 98.9508% | 98.9508% | 99.9212% |
| 4 | 8 | 9 | 7.5425 | 99.84% | 32.1717% | 89.2021% | 89.2021% | 99.2128% | 99.6805% | 99.9842% |
| 4 | 9 | 15 | 12.9708 | 100.00% | 25.3987% | 80.5595% | 99.6900% | 99.6900% | 99.9772% | 99.9952% |
| 4 | 10 | 9 | 8.1714 | 100.00% | 95.0478% | 95.0478% | 98.2804% | 98.2804% | 99.9421% | 99.9962% |

To benchmark our results against the greedy search algorithm of Section 5.iii, we calculated $APCS(x)$ for the 5 values of $x$ in table 2 when the original parameters were used. These evaluated to $APCS(99\%) = APCS(98\%) = APCS(97\%) = -82.00\%$, and $APCS(96\%) = APCS(95\%) = -19.76\%$. Recall that APCS proper for the greedy algorithm was -82.00%. In order to generate a bound greater than 95%, $x$ must be set to 81% or lower, effectively taking the (unreasonable) position of being ambivalent between utilities that are within 81% of the true optimum.